\documentstyle[12pt]{article}
\begin{document}
\vspace{72pt}
\begin{center}
{\large \bf
GRAND-CANONICAL ENSEMBLE OF RANDOM
 SURFACES WITH FOUR SPECIES OF ISING SPINS }\\
\vspace{36pt}
J.-P. KOWNACKI \footnote{Electronic mail:
kownacki@qcd.th.u-psud.fr} and
A. KRZYWICKI \footnote{Electronic
 mail: krz@qcd.th.u-psud.fr}\\
\vspace{10pt}
Laboratoire de Physique Th\' eorique
et Hautes Energies,\\
 B\^{a}t. 211, Universit\' e de Paris-Sud,
 91405 Orsay, France \\
\vspace{32pt}
{\bf Abstract }
\end{center}
The grand-canonical ensemble of dynamically
triangulated surfaces coupled to four species
of Ising spins (c=2) is simulated on a computer.
The effective string susceptibility exponent for
lattices with up to 1000 vertices is found to be
$\gamma = - 0.195(58)$. A specific scenario
for $c  > 1$ models is conjectured.
\vspace{32pt}

\noindent
PCAC number(s): 11.15.Ha, 11.17.+y, 64.60.Fr\\
\vspace{72pt}
\par\noindent
January 1994\\
LPTHE Orsay 94/11\\
\newpage
\section{INTRODUCTION}
The purpose of this paper is to present
a calculation of the string susceptibility
exponent for a statistical system of
dynamically triangulated random surfaces
coupled to four species of Ising spins.
The method employed consists in
simulating on a computer the so-called
grand-canonical ensemble of surfaces
with the topology of a sphere. The
motivation for this study is to contribute
to a better understanding of the physics
of random surfaces beyond the "barrier"
at $c=1$ ($c$ denotes the central charge
of matter fields interacting with the geometry).
\par
As is well known, the spectacular progress
in our understanding of two-dimensional
quantum gravity, achieved during the last
years, concerns models where $c \leq 1$.
The string susceptibility exponent
$\gamma$ is then given, for spherical
topology, by the formula

\begin{equation}
\gamma = {1 \over 12} [c - 1 - \sqrt{(c-1)(c-25)} \;]  ,
\label{kpz}
\end{equation}

\noindent
derived first, in the continuum framework,
by Knizhnik, Polyakov and
Za\-mo\-lod\-chi\-kov
\cite{kpz}. The right-hand
side of (\ref{kpz}) becomes complex for
$c > 1$, which is clearly unphysical and
indicates that something rather dramatic
happens when one attempts to move $c$
beyond unity. It is generally believed
that this "decease" reflects the tachyonic
instability of string models. A suggestive
heuristic picture imputes the
break-down of the
Euclidean theory for $c > 1$ to a
condensation of gravitational
singularities \cite{cat}.
\par
Eq. (\ref{kpz}) also holds
in solvable models of discrete quantum
gravity, when the "dynamical triangulation"
recipe \cite{dav} is adopted to discretize the
two-dimensional manifold.
Nevertheless, the {\em discrete} models of
Euclidean quantum gravity are well defined
for any value of $c$. Triangulated surfaces
collapse or degenerate into branched polymers
when $c \to \infty$ \cite{mig,adfo}. The
indications of the phenomenon have been
observed numerically, for
$c \; {\mbox{\lower0.6ex\hbox{\vbox{
\offinterlineskip \hbox{$>$}
\vskip1pt\hbox{$\sim$}}}}   } \; 10$, many
years ago \cite{adfo,jkp}.
Also, there exist theorems establishing the
 absence of a non-trivial continuum limit for
classes of discrete models (for a lucid discusion
of these issues we refer the reader to an
excellent review by Ambj\o rn \cite{ambrev}).
However, in numerical simulations, one does
not observe anything special to happen as
one crosses the famous "barrier". Moreover,
an analysis of a series expansion of the
partition function, rewritten using the
matrix model formalism, has been carried
out by Br\' ezin and Hikami \cite{brez}.
They also did not find any sign of pathology
at $c > 1$. This situation has motivated a
series of numerical simulations, with the
aim to achieve a better understanding of
the quantum gravity coupled to matter
with central charge larger than, but close
 to unity. As these studies were
 restricted to the microcanonical ensemble
 \cite{bj,kog,adjt,enzo}, we have decided
 to use the grand-canonical ensemble to
 determine directly the particularly
interesting susceptibility exponent, for
the model with four species of Ising
spins (c=2). In parallel, a novel method
of measuring this exponent has been
developed in Copenhagen \cite{thor1}.
When completing this work, we received
 the very recent ref. \cite{thor2}, where
$\gamma$ is also measured for the model
 we have been working with. Our calculation
 constitutes an independent check of these
results, obtained with a more traditional
method. In the last section of this paper
we also conjecture a specific scenario for
$c > 1$ models.

\section{THE NUMERICAL EXPERIMENT}
\subsection{The method}
The model is defined by the
partition function

\begin{equation}
Z(g,\beta) = \sum_{T,N} W(T) \;  e^{-g N}
\left( \sum_{[\sigma]} {\rm exp} [ \;
 \beta \sum_{<ij>} \sigma_i \sigma_j
\; ] \right)^n  \; ,
\label{partf}
\end{equation}

\noindent
where T refers to a specific triangulation
and $W(T)$ is the corresponding symmetry
factor. Further,  $N$ is the number of vertices
and $n$ is the number of species of Ising
spins $\sigma$. The cosmological
constant $g$ and the
spin coupling $\beta$ are the only parameters
of the model. Only nearest neighbour
spins are coupled. As in the early ref.
\cite{jkps} the spins live on the
vertices of the triangulated surface
(and not the dual one).
\par
The partition function can be rewritten as

\begin{equation}
Z=\sum_N e^{-g N} \; Z(N)
\label{partfN}
\end{equation}

\noindent
For large $N$, one expects

\begin{equation}
N^3 Z(N)\sim N^\gamma \; e^{g_{cr} N} \; ,
\label{as}
\end{equation}

\noindent
where $\gamma$ is the susceptibility
exponent and $g_{cr}$ is the (non-universal)
critical cosmological constant.
\par
A single step of the grand-canonical
algorithm \cite{adfo,jkp} changes $N$
by $\pm 1$. The code employed in this
 work to update the geometry is
essentially identical to the one used
and described in detail in ref. \cite{jkp}.
The only modification is that the
popular method due originally
to Baumann \cite{baum} is adopted: the partition
function is modified into

\begin{equation}
\tilde{Z} = \sum_N c(N) e^{-g N} N^3 Z(N) \; ,
\label{modf}
\end{equation}

\noindent
where $c(N) = c_l e^{\Delta N}$ for $N
\leq N_0$ and $c(N) = c_u e^{-\Delta N}$
for $N \geq N_0$ , while $\Delta$ is an
adjustable parameter. Of course, one has
$c_l / c_u = \exp{(-2 \Delta N_0)}$. Notice,
that the parameters $c_{l,u}$ need not be
specified individually since only their ratio
enters the detailed balance equations.
\par
Furthermore, $N$ is restricted to $N_0 \pm
 \delta N$, with $\delta N \ll N_0$.
Denote by $E(N)$ the experimentally measured
density. Since the algorithm satisfies the
detailed balance equations, within the finite
interval under consideration $E(N)$ tends to
the expression under the sum in (\ref{modf})
for infinite statistics, up to global normalization.
A sweep of geometry is defined as $N_0$
successive steps of the grand-canonical algorithm.
\par
Another difference, compared to ref. \cite{jkp}, is that
we consider Ising spins instead of Gaussian
fields. When a new vertex is created the new
spin(s) are generated using the heat bath
method. A sweep of geometry is always
followed by a sweep
of spins. The latter is performed using
Wolff's \cite{wolf} cluster algorithm.
This algorithm is particularly appropriate
in the vicinity of the critical
spin coupling $\beta_c$. Sufficiently above $\beta_c$
the clusters become large and the algorithm
loses its efficiency. For such values of
$\beta$ the heat bath algorithm becomes
competitive. We have used it occasionally
as an independent check of our results.
\par
Define

\begin{equation}
g_{eff}(n,N_0)={1 \over {2n}} {\rm ln}
\left({{E(N_0+n)} \over
{E(N_0-n)}}\right) \; + g
\label{geff}
\end{equation}

\noindent
Taking the asymptotic form (\ref{as}), one gets:

\begin{equation}
{ {E(N_0+n)} \over {E(N_0-n)}}
= e^{-2n(g-g_{cr})} \left({{N_0+n} \over
{N_0-n}} \right)^{\gamma}
\label{geff2}
\end{equation}

\noindent
and consequently

\begin{equation}
g_{eff}(n,N_0) = g_{cr} + {\gamma \over {N_0}}
\label{geff3}
\end{equation}

\noindent
to leading order in $N_0$. The quantity
$g_{eff}(n,N_0)$ is measured for a set of
values of  $N_0$. Since the right-hand side of
(\ref{geff3}) is independent of $n$, one
averages over $n$ to get an
improved estimate of $g_{eff}(N_0)$. The
procedure is repeated many times
and  the error on $g_{eff}(N_0)$
is found with the binning method.
The parameters $\gamma$
and $g_{cr}$ are then determined
from the linear fit to $g_{eff}$
versus $N_0^{-1}$.
\par
The choice of $\Delta$ is a pure
matter of convenience. The experimental
 histogram $E(N)$ should fall
exponentially as one moves off the
point $N = N_0$, but one wants to have
 good statistics for $n \leq 5$, say, and
therefore this fall should not be too fast.
 The choice of the input value of $g$
involves a subtlety. For large $N_0$
one has roughly

\begin{equation}
{{E(N_0 +n )} \over {E(N_0-n)}} \sim
e^{-2n (g-g_{cr})}
\label{rough}
\end{equation}

\noindent
Thus, the experimental histogram is
asymmetric. This asymmetry, together
 with the statistical fluctuations of
the histogram, generates a systematic
 error in the measurement of $g_{eff}(n, N_0)$.
For example, if for a given $n$ one
has $Z_{exp}(N_{0}+n) < Z_{exp}(N_{0}-n)$,
 then what is measured is in fact
$g_{eff}(n,N_{0})-\delta$, where
$\delta \sim (1/4n)(\epsilon_n^2-
\epsilon_{-n}^2 )$  and
$\epsilon_{\pm n}$ is the statistical
error on $E(N_0 \pm n)$. In practice,
one has to start with some exploratory
runs in order to get a first estimate of
$g_{cr}$, and then use this estimate as the
input $g$ in the full scale simulation.
Proceeding in this way one makes the
systematic error insignificant
compared to the statistical one.

\subsection {Simulating solvable models}
In order to check and gauge our "tools"
two solvable models are simulated first,
viz. pure gravity and the model with 1 spin/site.
\par
{\em Pure gravity $(c = 0)$}: The
string  susceptibility exponent $\gamma$ is
known to be $-1/2$. The expression
for the partition function is exactly known and
the critical value of the fugacity $z \equiv \exp{(-g)}$ is
$z_{cr}= 27/ 256 \approx  0.1055$.
To check the code, it is verified that
$E(N)$, once normalized, follows very closely
the exact formula. Then, $z$ is set to value
$0.1044$, on purpose slightly off the critical
one, and we perform experiments at
$N_0$=100, 120, 150, 200, 300, 500 and 1000.
The Baumann parameter is given the
value $\Delta = \ln{2}$, forcing $E(N)$ to
decrease by about a factor 5 between $N_0$ and
$N_0 \pm 10$. The number of sweeps per
experiment is about 1 to 4$\times10^5$
sweeps (following 4000 heating sweeps) and
the observed acceptance of the
grand-canonical algorithm
is found to be about
72$\%$ independently of $N_0$.
\par
This exercise is instructive, as it helps
appreciating the influence of finite size
corrections on the results of an experiment
of this type. Using all data points the
estimate $\gamma = -0.563(33)$ is produced,
significantly different from the expected
$-0.5$ . This is not just a fluctuation: the
effective exponent is below $-0.5$
precisely because our data follow closely
the exact formula. However, without the points at
$N < 200$ one obtains $\gamma =
-0.509(56)$, with a bigger error (there are
less data) but closer to
the theoretical value. Of course, in
this particular model one can determine
analytically the threshold for the
applicability of (\ref{as}) and the form
of correction terms. This is not the
case for more complicated models.
We believe that, while interpreting
data obtained in simulating  such
models, guessing the corrections
 to (\ref{as}) is a waste of time. It
is preferable to stick to the simple
expression (\ref{as}) and to accept
 the fact of life that one can only
get an {\em effective} exponent.
 Anyway, this is the philosophy we
have adopted.
\par
{\em 1 spin/site ($c = 1/2)$}:
The string susceptibility
 is now $\gamma = - 1/3$.
The critical spin coupling is also known exactly:
$\beta_{cr}=-{1\over 2}
\ln{\tanh{{1 \over 2}\ln{(108/23)}}} \; \approx \; 0.2163$ \cite{bur}.
The code is tested comparing $E(N)$ to the
first few terms of the expansion in powers
of $z$ calculated analytically; the expected values
are reproduced with an
error $ < 0.1 \%$. We set
 $\Delta = \ln{2}$ , $\delta N =
5$ and $\beta= \beta_{cr}$.
The value of fugacity $z = 0.05$ is found to be
acceptable as the input one.
One sweep is defined as $N_0$ steps of
the grand-canonical algorithm followed by
$n_w =  N / \langle {\rm cluster \;  size}\rangle$
calls to the Wolff algorithm. The
average  size of clusters produced by the
Wolff algorithm is determined
prior to each production run and it is
checked that is remains stable
during the run. The efficiency of the
grand-canonical algorithm is about
$57\%$. It is checked, using either a "cold"
or  a "hot" start, that the system thermalizes
 after about 2000 sweeps (cf. ref \cite{jkps}).
This is therefore the number of sweeps to be
performed initially in order to heat the
 system.  Simulations are carried out for
$N_0$ = 200, 300, 500 and 1000 with
about $1.3 \times 10^5$ sweeps
for each $N_0$. The resulting exponent is
$\gamma=-0.317(102)$. This is unprecise
but correct and, without further ado, we go
over to the truly interesting case of 4 spins/site.

{\subsection{Surfaces with 4 spins/site}
Now, neither the string susceptibility
exponent, nor the critical spin coupling
$\beta_{c}$ are known theoretically. Of
course, it is expected that
$\gamma(\beta)$ has a maximum at
$\beta=\beta_{c}$ and falls off to $- 1/2$
as $\beta$ moves away from its critical
value.  The exponent $\gamma$ is estimated for
several values of $\beta$ in order to
observe the maximum in question and
the corresponding value of $\gamma$.
We proceed as in the 1 spin/site
case, making the same tests (with similar
results) and choosing the same
values of $\Delta$ and $\delta N$.
The fugacity is always chosen so that $E(N)$
is almost symmetric in the neighborhood of $N_0$.
The observed efficiency of the grand-canonical
 algorithm is about 56$\%$ .
The experiment is carried out for $N_0$=200,
 300, 500 and 1000, with
2000 heating sweeps followed
by $ 1.3 \times 10^6$ production
sweeps, for each $N_{0}$.
\par
The results are shown in fig. 1. The exponent
 $\gamma$ takes its maximum value near
$\beta \approx 0.196$. This is reassuring,
since this is the critical coupling found from
microcanonical simulation in ref. \cite{bj}
(in comparing with their value remember
that they have been working on the dual
lattice).
\par
The relatively slow fall of $\gamma$
on the right of the critical point is associated
with the worsening performances of the Wolff
method (size of clusters increasing too much). We have
also made some runs with the heat bath algorithm.
With the latter, in order to get results close to
those with Wolff, one has (for the system with
4 spins/site) carry out 3 sweeps of spins
after each sweep of the geometry. And, as
expected, the autocorrelation time is longer
(however not larger than $\sim 100$ sweeps, this
low number is presumably due to the fact that
the grand-canonical algorithm permanently
destroys and creates spins). But at $\beta =
0.24$ the value obtained using the heat bath
algorithm is $-0.487(83)$, which is closer to
the expected $-1/2$ than the rightmost point in fig. 1.

\section{DISCUSSION AND SPECULATIONS}

Using the two top values from fig. 1 we
get the estimate $\gamma=-0.195(58)$. It is
compatible with the value found in ref.
\cite{thor2}, which is $-0.167(8)$. The
respective errors should not be directly
 compared, since they have different
 significance. Our error estimate is very
 conservative. The error given in ref.
 \cite{thor2} is the error obtained in
fitting the baby universe distribution
in a single experiment. Contemplating
the dispersion of  $\gamma$'s  shown
in fig. 1 of ref. \cite{thor2} and corresponding
 to nearby values of $\beta$, one suspects
that this error should be multiplied by at
 least a factor of 2. Nevertheless, our
experiment is admittedly less precise,
simply due to our lower statistics.
\par
The size of our lattices is the same
as the size of baby universes
observed in ref. \cite{thor2}.
However, with their method one
can easier extend simulations to
 larger systems, and it is this which
 makes it superior to the grand-canonical
technique employed in our work.
Anyhow, it is certainly a good thing
to have two independent and
compatible estimates of the string
susceptibility exponent at $c = 2$,
 obtained with two completely different methods.
\par
We wish to end this paper with a speculation
concerning  $c > 1$ models. We conjecture that
the continuum $c = 1$ theory acts as an
"attractor" within a finite interval
$c_0 > c \geq 1$. More precisely, for these
 central charges, we suspect the {\em long
distance} physics of the different models to
 be the same, modulo a renormalization of
the length scale. In particular, this means
that the microcanonical partition function
behaves as follows:

\begin{equation}
N^3 Z(N,c) \sim {{e^{g_{cr} N}} \over
 {\ln^2{(\lambda N)}}} \; [ 1 + ...]   \;  ,
\label{spec}
\end{equation}

\noindent
where $\lambda = \lambda(c)$ is some
constant and the dots represent subleading
non-universal contributions. One can
check that the normalized distribution of
 "baby universes" given in ref. \cite{thor2}
for $c = 2$ is compatible with (\ref{spec}),
 even without sub-leading corrections,
provided one sets $\lambda \approx 20$.
 With such a value of $\lambda$ the effective
 $\gamma$ calculated from (\ref{spec}) for
 $N \in [200, 1000]$ is also close to the one
 measured in this work. Such a
 simple rescaling does not reproduce the data
 \cite{thor1,thor2} for $ c \geq 3$. Of course,
 one could reconcile (\ref{spec}) with these
data playing with the sub-leading terms.
This would be a futile exercise, however,
in view of the lack of any theoretical
prediction about the latter.
\par
Our conjecture is simple and refutable. We
obviously expect that the effective
susceptibility exponent tends to zero when
data with larger $N$ are used, provided
 $1 < c < c_0$. Actually, the authors of ref.
 \cite{thor2} report that for $c = 1.6$, an
extension by the factor of 10 of the size
 of "baby universes"  produces a shift of
the effective $\gamma$ from $- 0.212(15)$
 to  $- 0.175(24)$. Because of the errors, the
 effect is not very significant. However, it
 goes in the right direction and the magnitude
 of the shift, if taken at the face value, is
just what we expect with $\lambda = 10$ to
 20. This is perhaps not an accident. If true,
 our conjecture also explains why the critical
exponents (fractal dimension and magnetic
 exponents) found in microcanonical simulations
 \cite{bj,kog,adjt,enzo} show so little
dependence on $c$ just beyond the "barrier".
\par
The general arguments reviewed in ref.
 \cite{ambrev} indicate that $\gamma > 0$
implies that this exponent has a finite value,
which for spin models is most likely ${1 \over 3}$
(see also ref. \cite{wex}).
We thus expect that, in the continuum limit,
the exponent $\gamma$ jumps from the value
0 it takes in the interval $[1, c_0)$ to a moderate
value, say ${1 \over 3}$ , for
$c \geq c_0$.  We imagine that
the jump occurs when the
density of baby universes increases above
some critical value. With this perspective, we are
tempted to consider the small  positive values of
$\gamma$ (e.g. $\gamma \approx 0.04$
to 0.06 for $c = 2$) extracted from series
expansion in ref. \cite{brez} as supporting
our conjecture.
\par
It has been suggested long time ago (first
by David in ref. \cite{dav}; see also
\cite{krz} and references therein)
that the effective $\gamma$ becomes
positive somewhere in the neighborhood
of $c = 4$. The most recent data
presented in ref. \cite{thor2} indicate
that this happens for $ 3 \; {\mbox
{\lower0.6ex\hbox{\vbox{
\offinterlineskip \hbox{$<$}
\vskip1pt\hbox{$\sim$}}}}   } \; c < 4$
This result suggests the crude estimate
$ 3 \; {\mbox{\lower0.6ex\hbox{\vbox{
\offinterlineskip \hbox{$<$}
\vskip1pt\hbox{$\sim$}}}}   } \; c_0 < 4$.
It is perhaps not unreasonable to think that
such a value of $c_0$ is large enough to
 warrant the approximate validity of the
mean field calculations for $c > c_0$.
\par
Although our conjecture differs from that
put forward in ref. \cite{thor2} as a plausible
"$c > 1$ hypothesis", it is likewise pessimistic:
in both scenarios the physics of a $c > 1$
model is reducible
to the known theories with $c \leq 1$.
\par
This study has been triggered by a
correspondence with D. Johnston, whom
we wish also to thank for interesting
conversations. We wish to thank J. Ambj\o rn
for useful correspondence. One of us (JPK) is indebted
to the French-British collaboration fund
 Alliance for covering expenses associated
with his visit at the Heriot-Watt University.
 Laboratoire de Physique Th\' eorique et
 Hautes Energies is Laboratoire associ\' e au C.N.R.S.

\vspace{50pt}
\noindent
{\Large {\bf Figure caption}}\\
\vspace{25pt}
\begin{description}
\item[Fig. 1] - The string susceptibility exponent
$\gamma$ versus the spin coupling $\beta$.\\
The arrow indicates the critical
 coupling calculated in ref. \cite{bj}.
\end{description}
\end{document}